# Generalised linear mixed model analysis via sequential Monte Carlo sampling

**Y. Fan**

*School of Mathematics and Statistics,*
*University of New South Wales, Sydney 2052, Australia*
*e-mail:* Y.Fan@unsw.edu.au

**D.S. Leslie**

*School of Mathematics,*
*University of Bristol, University Walk,*
*Bristol, BS8 1TW, United Kingdom*
*e-mail:* David.Leslie@bristol.ac.uk

**M.P. Wand**

*School of Mathematics and Applied Statistics,*
*University of Wollongong, Wollongong 2522, Australia*
*e-mail:* mwand@uow.edu.au

**Abstract:** We present a sequential Monte Carlo sampler algorithm for the Bayesian analysis of generalised linear mixed models (GLMMs). These models support a variety of interesting regression-type analyses, but performing inference is often extremely difficult, even when using the Bayesian approach combined with Markov chain Monte Carlo (MCMC). The Sequential Monte Carlo sampler (SMC) is a new and general method for producing samples from posterior distributions. In this article we demonstrate use of the SMC method for performing inference for GLMMs. We demonstrate the effectiveness of the method on both simulated and real data, and find that sequential Monte Carlo is a competitive alternative to the available MCMC techniques.

## 1. Introduction

Effective strategies for generalised linear mixed model (GLMM) analysis continues to be a vibrant research area. Reasons include:

- GLMMs have become an indispensable vehicle for analysing a significant portion of contemporary complex data sets.
- GLMMs are inherently difficult to fit compared with ordinary linear mixed models and generalised linear models.

- Existing strategies involve a number of trade-offs concerning, for example, approximation accuracy, computational times and Markov chain convergence.

Overviews of the usefulness and difficulties of GLMM-based analysis may be found in, for example, [23, 27] and [28].

Most practical GLMM methodology falls into two categories: analytic approximations (e.g. [2]) and Monte Carlo methods (e.g. [6]). Monte Carlo methods have the advantage of providing direct approximations to quantities of interest [1]. On the other hand, analytic approximations, such as Laplace approximation, are indirect and prone to substantial bias (e.g. [3]). The most common Monte Carlo approach is Markov Chain Monte Carlo (MCMC), where approximation accuracy is associated with Markov chain convergence.

[30] is a recent example of research concerned with practical GLMM analysis via Markov chain Monte Carlo. Those authors explored use of the MCMC computing package `WinBUGS` and showed it to exhibit good performance for a number of examples.

One of the major difficulties associated with using MCMC is the need to assess convergence. Popular methods for convergence assessment often rely on the comparison of multiple sample output; see [7] for a comparative review. These methods can invariably fail to detect a lack of convergence and one needs to be cautious when taking such an approach. Another major drawback of MCMC is the difficulty in designing efficient samplers for complex problems. The use of historical information from MCMC sample paths has to be treated very carefully, so that the equilibrium distribution of the Markov chain is not disturbed. Various methods have been proposed in the literature, (see [14]), however the practical applicability of these so-called adaptive methods can be limited.

Both problems associated with MCMC discussed above are inherently due to the Markovian nature of the MCMC sampler. Sequential Monte Carlo (SMC) methods provide an alternative framework for posterior sampling, which is not dependent on the convergence of a Markov chain as in the MCMC sampler case. Though careful assessment of posterior samples is also applicable in the SMC case, these are more in line with the standard Monte Carlo methods. SMC samplers can be seen as an extension of the well known importance sampling method. The fact that SMC methods do not rely on Markov chain theory means that it is a more flexible sampler. In the sense that for example, if the historical sample path of the SMC sampler is informative for the design of an efficient algorithm, this can be done quite easily in the SMC framework. In this article we show that sequential Monte Carlo methods provide an effective means of Bayesian GLMM analysis. We provide a general yet simple framework for efficient design of the sampler, and demonstrate that this approach is a viable alternative to MCMC, and since SMC samplers require a number of user-specified inputs, we will give recommendations in the GLMM framework on how these are chosen.

Section 2 contains a brief summary of Bayesian approaches to generalised linear mixed models. In Section 3 we provide details on analysis for such models via sequential Monte Carlo sampling. In Section 4 we present two examples. In a simulated Poisson regression example, we compare the efficiencies of the SMC sampler with alternative Monte Carlo methods, and then demonstrate the effectiveness of the SMC sampler in a binary logistic regression example involving real data. Some comparisons of algorithm efficiencies for the two examples are carried out in Section 5 and concluding remarks are given in Section 6. The software used for this paper is available from the authors on request.

## 2. Bayesian generalised linear mixed models

GLMMs for canonical one-parameter exponential families (e.g. Poisson, logistic) and Gaussian random effects take the general form

$$[\mathbf{y}|\boldsymbol{\beta}, \mathbf{u}, \mathbf{G}] = \exp\{\mathbf{y}^T(\mathbf{X}\boldsymbol{\beta} + \mathbf{Z}\mathbf{u}) - \mathbf{1}^T b(\mathbf{X}\boldsymbol{\beta} + \mathbf{Z}\mathbf{u}) + \mathbf{1}^T c(\mathbf{y})\}, \qquad (1)$$

$$[\mathbf{u}|\mathbf{G}] \sim N(\mathbf{0}, \mathbf{G}) \qquad (2)$$

where here, and throughout, the distribution of a random vector $\mathbf{x}$ is denoted by $[\mathbf{x}]$ and the conditional distribution of $\mathbf{y}$ given $\mathbf{x}$ is denoted by $[\mathbf{y}|\mathbf{x}]$. In the Poisson case $b(x) = e^x$, while in the logistic case $b(x) = \log(1 + e^x)$. An important special case of (1)–(2) is the variance components model

$$[\mathbf{y}|\boldsymbol{\beta}, \mathbf{u}, \sigma_{u1}^2, \ldots, \sigma_{uL}^2] = \exp\{\mathbf{y}^T(\mathbf{X}\boldsymbol{\beta} + \mathbf{Z}\mathbf{u}) - \mathbf{1}^T b(\mathbf{X}\boldsymbol{\beta} + \mathbf{Z}\mathbf{u}) + \mathbf{1}^T c(\mathbf{y})\},$$

$$\mathbf{u} = \begin{bmatrix} \mathbf{u}_1 \\ \vdots \\ \mathbf{u}_L \end{bmatrix},$$

$$[\mathbf{u}|\sigma_{u1}^2, \ldots, \sigma_{uL}^2] \sim N(\mathbf{0}, \text{blockdiag}_{1 \le \ell \le L}(\sigma_{u\ell}^2 \mathbf{I}_{q_\ell})). \qquad (3)$$

where $q_\ell$ is the number of elements in $\mathbf{u}_\ell$. While (3) is not as general as (1)–(2) it still handles many important situations such as random intercepts and generalised additive models [30]. With simplicity in mind, we will focus on this GLMM for the remainder of the paper. However, the methodology of Section 3 is quite general and is extendable to more elaborate GLMMs.

In this study we have worked with diffuse conjugate priors although, once again, the methodology extends to other types of priors. To ensure scale-invariance all continuous predictors are standardised at the start of the Bayesian analysis. The prior on $\boldsymbol{\beta}$ is a diffuse Gaussian:

$$\boldsymbol{\beta} \sim N(\mathbf{0}, \sigma_\beta^2 \mathbf{I}) \qquad (4)$$

for some large $\sigma_\beta^2 > 0$. The prior for $(\sigma_{u1}^2, \ldots, \sigma_{uL}^2)$ is assumed to have independent components; i.e.

$$[\sigma_{u1}^2, \ldots, \sigma_{uL}^2] = [\sigma_{u1}^2] \cdots [\sigma_{uL}^2].$$

A number of possibilities for $[\sigma_{u\ell}^2]$ could be considered [18]. These include an inverse gamma distribution, a uniform distribution, and a folded Cauchy distribution. In this paper we use a conditionally conjugate inverse gamma distribution:

$$[\sigma_{u\ell}^2] = \frac{A_{u\ell}^{A_{u\ell}}}{\Gamma(A_{u\ell})}(\sigma_{u\ell}^2)^{-A_{u\ell}-1} e^{-A_{u\ell}/\sigma_{u\ell}^2}, \quad \sigma_{u\ell}^2 > 0 \ . \qquad (5)$$

This prior distribution was advocated by [30] for $A_{u\ell} = 0.01$. The prior is therefore fairly non-informative, yet results in a slightly simpler sampling procedure.

It will be convenient to introduce some additional notation to enable the model to be described more succinctly. We start by writing

$$\mathbf{C} = [\mathbf{X} \ \mathbf{Z}] \quad \text{and} \quad \boldsymbol{\nu} = \left[ \begin{array}{c} \boldsymbol{\beta} \\ \mathbf{u} \end{array} \right].$$

We also write $q_\beta$ for the number of elements in $\beta$, and

$$\mathbf{V} = \text{blockdiag}(\sigma_\beta^2 \mathbf{I}_{q_\beta}, \sigma_{u1}^2 \mathbf{I}_{q_1}, \ldots, \sigma_{uL}^2 \mathbf{I}_{q_L})$$

for the prior covariance of $\boldsymbol{\nu}$. Writing $\boldsymbol{\sigma}^2$ for $(\sigma_{u1}^2, \ldots, \sigma_{uL}^2)$, we can then combine (3), (4) and (5) to give the joint density of all parameters and data:

$$[\mathbf{y}, \boldsymbol{\nu}, \boldsymbol{\sigma}^2] = \exp\left[ \mathbf{y}^T \mathbf{C}\boldsymbol{\nu} - \mathbf{1}^T b(\mathbf{C}\boldsymbol{\nu}) + \mathbf{1}^T c(\mathbf{y}) - \tfrac{1}{2}\boldsymbol{\nu}^T \mathbf{V}^{-1}\boldsymbol{\nu} - \sum_{\ell=1}^{L} \frac{q_\ell}{2} \log(\sigma_{u\ell}^2) \right.$$
$$\left. + \sum_{\ell=1}^{L} \left\{ A_{u\ell} \log(A_{u\ell}) - \log\Gamma(A_{u\ell}) - (A_{u\ell}+1)\log(\sigma_{u\ell}^2) - A_{u\ell}/\sigma_{u\ell}^2 \right\} \right].$$

From this, and noting that $\boldsymbol{\nu}^T \mathbf{V}^{-1}\boldsymbol{\nu} = \|\boldsymbol{\beta}\|^2/\sigma_\beta^2 + \sum_{\ell=1}^{L} \|\mathbf{u}_\ell\|^2/\sigma_{u\ell}^2$, it is clear that the density of the posterior distribution of the parameters is simply proportional to the function

$$\pi(\boldsymbol{\nu}, \boldsymbol{\sigma}^2) = \exp\left[ \mathbf{y}^T \mathbf{C}\boldsymbol{\nu} - \mathbf{1}^T b(\mathbf{C}\boldsymbol{\nu}) - \frac{1}{2\sigma_\beta^2}\|\boldsymbol{\beta}\|^2 \right.$$
$$\left. - \sum_{\ell=1}^{L} \left\{ (A_{u\ell} + \tfrac{q_\ell}{2} + 1)\log(\sigma_{u\ell}^2) + (A_{u\ell} + \tfrac{1}{2}\|\mathbf{u}_\ell\|^2)/\sigma_{u\ell}^2 \right\} \right]. \quad (6)$$

In Section 3, we will develop a sequential Monte Carlo sampler to produce samples from the distribution proportional to $\pi$.

## 3. Sequential Monte Carlo sampling

The Monte Carlo approach to GLMM analysis performs inference by drawing samples from the joint posterior distribution of the parameters $\boldsymbol{\theta} = (\boldsymbol{\beta}, \mathbf{u}, \sigma_{u1}^2, \ldots, \sigma_{uL}^2)$. We write $\pi(\boldsymbol{\theta})$ for the (unnormalised) density of this posterior distribution. Instead of using a Markov chain with $\pi$ as its stationary distribution to produce these samples, the sequential Monte Carlo (SMC) sampling method is a generalisation of importance sampling that produces a weighted sample from $\pi$ while retaining some of the benefits of MCMC analysis [8].

The use of SMC for static problems (as opposed to particle filters for dynamic problems; [12] requires the introduction of auxiliary distributions $\pi_0, \pi_1, \ldots, \pi_{S-1}$. At stage $s$ of the sampler we use a (weighted) sample from the previous distribution $\pi_{s-1}$ to produce a (weighted) sample from $\pi_s$. We set $\pi_S = \pi$ so that after $S$ stages we have a sample from the posterior distribution of interest. The introduction of the intermediate distributions allows the initial distribution $\pi_0$ to be gradually corrected to resemble the target distribution $\pi$, and can often overcome problems such as particle depletion where, if the two consecutive distributions are too dissimilar, then a small number of particles carry all the weight in the final sample.

The auxiliary distributions can be constructed in several ways: [5] introduces the observations incrementally to evolve the distribution from the prior to the posterior; [13] uses a similar technique, but increases the size of the state space as more observations are added; [8] use

$$
\begin{aligned}
\pi_s &\propto \pi_0^{1-\gamma_s} \pi^{\gamma_s}, \quad \text{where} \\
0 &= \gamma_0 \le \gamma_1 \le \cdots \le \gamma_S = 1
\end{aligned}
\tag{7}
$$

and $\pi_0$ is chosen to be the prior distribution for the parameters. In this article, due to the diffuse nature of the prior distribution, the initial distribution $\pi_0$ is instead chosen to be a multivariate Normal distribution with mean and covariance matrix chosen based on estimates obtained using classical methods for fitting GLMMs.

The SMC sampler algorithm starts by sampling $N$ samples, termed "particles", from the initial distribution $\pi_0$. Denote by $\boldsymbol{\theta}_i^0$ the $i$th particle at initial stage $s = 0$, and allocate weight $w_i^0 \equiv 1$ to each of the $N$ particles, so that $\{\boldsymbol{\theta}_i^0, w_i^0\}$ is a weighted sample from $\pi_0$.

The SMC sampling technique uses the weighted particles from distribution $\pi_{s-1}$ to produce particles from distribution $\pi_s$ through moving, reweighting and (possibly) resampling; see [8]. For simplicity, the formulation we use is that described in detail in Section 3.3.2.3 of that paper, which essentially results in the resample–move algorithm used by [5] and [19]. This is also similar to the annealed importance sampling method of [24], but the use of resampling within the algorithm greatly improves the efficiency of the method. Writing $\boldsymbol{\theta}_i^s$ for the $i$th particle at stage $s$, at each stage $0 < s \le S$ of the algorithm we perform the following steps:

**Reweight** Given $N$ weighted particles $\{\boldsymbol{\theta}_i^{s-1}, w_i^{s-1}\}$ from $\pi_{s-1}$, set

$$
w_i^s = w_i^{s-1} \frac{\pi_s(\boldsymbol{\theta}_i^{s-1})}{\pi_{s-1}(\boldsymbol{\theta}_i^{s-1})}.
$$

$\{\boldsymbol{\theta}_i^{s-1}, w_i^s\}$ is now a weighted sample from $\pi_s$.

**Resample** If the effective sample size (ESS, [20]), defined as $(\sum_{i=1}^N w_i^s)^2 / \sum_{i=1}^N (w_i^s)^2$, is less than $kN$, where $k$ is some constant typically taken to be $1/2$, then we perform stratified resampling [21]. ESS estimates the number of simple random samples from the target distribution that are required to obtain an estimate with the same Monte Carlo variation as the estimate using the $N$ weighted particles. Resampling refers to a suite of techniques that replicate the particles in such a way that the expected value of particle-based estimators is retained, but particles with low weights are discarded and particles with high weights multiplied; see [11] for a summary and comparison of several such approaches. This standard practise in the sequential Monte Carlo literature allows further computational effort to focus on samples that are likely to contribute non-negligibly to the final estimate. Finally, resampled particle weights are reset to $\{w_i^s\} \equiv 1$.

**Move** Let $\{\tilde{\boldsymbol{\theta}}_s, w_i^s\}, i = 1, \ldots, N$ denote samples from at the current distribution $\pi_s$ after reweighting and (possibly) resampling. To increase particle diversity we replace each sample according to

$$\boldsymbol{\theta}_i^s \sim K_s(\tilde{\boldsymbol{\theta}}^s{}_i, \cdot)$$

where $K_s$ is an MCMC transition kernel that admits $\pi_s$ as stationary distribution. [15] provides detail on MCMC transition kernels.

It is known [8] that this particular formulation of the SMC sampling is suboptimal, in terms of the variance of the importance weights $\{w_i^s\}$, especially if the distributions on consecutive stages are too far apart. However, since the optimal formulation is intractable, and for the static problem we have here it is easy to ensure that the difference between $\pi_{s-1}$ and $\pi_s$ is small, we use this simpler formulation. (Contrast this situation with that of an SMC algorithm for a dynamic problem, or the technique of [5] where data arrive over time and there is no control over the distance between $\pi_{s-1}$ and $\pi_s$.)

The "parameters" of the algorithm that must be chosen when implementing this sampler are therefore:

- the initial distribution $\pi_0$,
- the sequence of values $\gamma_s$ that govern the rate of transition from the initial distribution $\pi_0$ to the posterior distribution $\pi$,
- the transition kernels $K_s$, used to move the particles within the distribution proportional to $\pi_s$, and
- the number of particles $N$.
- the total number of distributions $S$.

Specific choices of these parameters used in this paper are discussed in the following subsections. We give a more algorithmic description of our method in the Appendix.

### 3.1. Initial distribution $\pi_0$

As previously observed, using the prior distribution as an initial distribution is flawed in this case, since the prior is highly diffuse. Instead we use the penalised quasi-likelihood (PQL) method [2] to obtain an approximate fit of the model. Let $\widehat{\boldsymbol{\nu}}_{\text{PQL}}$ and $\widehat{\boldsymbol{\sigma}}^2_{\text{PQL}}$ be the estimate of $\boldsymbol{\nu}$ and $\boldsymbol{\sigma}^2$ obtained using PQL. We will calculate a normal approximation of the posterior distribution of $\boldsymbol{\nu}$ centred at this approximate maximum likelihood estimate, which can then be used to construct an initial distribution $\pi_0$ for the SMC sampling procedure. Note from

(6) that

$$\pi(\boldsymbol{\nu}, \boldsymbol{\sigma}^2) = \exp\left\{\mathbf{y}^T\mathbf{C}\boldsymbol{\nu} - \mathbf{1}^T b(\mathbf{C}\boldsymbol{\nu}) - \tfrac{1}{2}\boldsymbol{\nu}^T\mathbf{V}^{-1}\boldsymbol{\nu} + f(\boldsymbol{\sigma}^2)\right\},$$

where $f$ is some function that does not depend on $\boldsymbol{\nu}$. It is a simple calculation to see that the matrix of second derivatives with respect to components of $\boldsymbol{\nu}$ is $-\mathbf{C}^T\mathrm{diag}\{b''(\mathbf{C}\boldsymbol{\nu})\}\mathbf{C} - \mathbf{V}^{-1}$; we therefore initialise our algorithm by taking a normal distribution for $\boldsymbol{\nu}$ with mean $\widehat{\boldsymbol{\nu}}_{\mathrm{PQL}}$ and covariance matrix

$$\boldsymbol{\Sigma} = \left[\mathbf{C}^T\mathrm{diag}\left\{b''(\mathbf{C}\widehat{\boldsymbol{\nu}}_{\mathrm{PQL}})\right\}\mathbf{C} + \widehat{\mathbf{V}}_{\mathrm{PQL}}^{-1}\right]^{-1}, \tag{8}$$

where the entries in $\widehat{\mathbf{V}}_{\mathrm{PQL}}$ are taken from $\widehat{\boldsymbol{\sigma}}_{\mathrm{PQL}}^2$.

It remains to specify a distribution for the variance vector $\boldsymbol{\sigma}^2$. We have found it convenient to specify this conditional on $\boldsymbol{\nu}$, and of a form that is consistent with the posterior distribution $\pi$. We take

$$\pi_0(\boldsymbol{\sigma}^2 \mid \boldsymbol{\nu}) = \prod_{\ell=1}^{L} \frac{(A_{u\ell} + \tfrac{1}{2}\|\mathbf{u}_\ell\|^2)^{(A_{u\ell} + q_\ell/2)}}{\Gamma(A_{u\ell} + q_\ell/2)}(\sigma_{u\ell}^2)^{-A_{u\ell} - q_\ell/2 - 1}e^{-(A_{u\ell} + \tfrac{1}{2}\|\mathbf{u}_\ell\|^2)/\sigma_{u\ell}^2}, \tag{9}$$

i.e. the $\sigma_{u\ell}^2$ are conditionally independent given $\boldsymbol{\nu}$, and each has an inverse gamma distribution depending on the corresponding components of $\mathbf{u}$. Hence, an initial sample from $\pi_0$ can easily be generated by first sampling from the normal distribution for $\boldsymbol{\nu}$ then sampling the $\sigma_{u\ell}^2$ from their conditional distributions. Furthermore, we will see in Sections 3.2 and 3.3 that this results in simple conditional distributions for $\sigma_{u\ell}^2$ at all stages of the sampler.

Putting together the initial distributions of $\boldsymbol{\nu}$ and $\boldsymbol{\sigma}^2$, we see that

$$\begin{aligned}
\pi_0(\boldsymbol{\nu}, \boldsymbol{\sigma}^2) \propto \exp\Bigg[ &-\frac{1}{2}(\boldsymbol{\nu} - \widehat{\boldsymbol{\nu}}_{\mathrm{PQL}})^T\boldsymbol{\Sigma}^{-1}(\boldsymbol{\nu} - \widehat{\boldsymbol{\nu}}_{\mathrm{PQL}}) \\
&+ \sum_{\ell=1}^{L}\bigg\{(A_{u\ell} + \frac{q_\ell}{2})\log(A_{u\ell} + \frac{1}{2}\|\mathbf{u}_\ell\|^2) \\
&- (A_{u\ell} + \frac{q_\ell}{2} + 1)\log\sigma_{u\ell}^2 - (A_{u\ell} + \frac{1}{2}\|\mathbf{u}_\ell\|^2)/\sigma_{u\ell}^2\bigg\}\Bigg]. \tag{10}
\end{aligned}$$

In the GLMM examples of this paper, there is no reason to suspect that the posterior distribution is particularly spread out or multi-modal. Hence this $\pi_0$ is sufficient, as demonstrated by the fact that several different Monte Carlo methods provide identical inference in the examples of Section 4. In other examples where these complications are likely to occur, $\pi_0$ should be chosen with an inflated variance, or to be a $t$ distribution, to help dominate the posterior distribution. Note that multi-modality is less of a problem for the sequential Monte Carlo approach than it would be for MCMC, since there is no difficulty in having samples in both modes simultaneously, whereas an MCMC approach must move between the nodes through areas of low posterior probability.

### 3.2. Sequence of intermediary distributions

In this section we describe the sequence of distributions used to transition from $\pi_0$ to $\pi_S = \pi$. Recall that we choose to use the formulation (7). Using (10) and (6) it is clear that the intermediate distributions are proportional to $\pi_s$ where

$$
\pi_s(\boldsymbol{\nu}, \boldsymbol{\sigma}^2)
$$

$$
= \exp\Big[\gamma_s\big\{\mathbf{y}^T\mathbf{C}\boldsymbol{\nu} - \mathbf{1}^T b(\mathbf{C}\boldsymbol{\nu}) - \frac{1}{2\sigma_\beta^2}\|\boldsymbol{\beta}\|^2 + \sum_{\ell=1}^{L}(A_{u\ell} + \frac{q_\ell}{2})\log(A_{u\ell} + \frac{1}{2}\|\mathbf{u}_\ell\|^2)
$$

$$
- \sum_{\ell=1}^{L}\big((A_{u\ell} + \frac{q_\ell}{2} + 1)\log(\sigma_{u\ell}^2) + (A_{u\ell} + \|\mathbf{u}_\ell\|^2)/\sigma_{u\ell}^2\big)\big\}
$$

$$
+ (1 - \gamma_s)\big\{-\frac{1}{2}(\boldsymbol{\nu} - \widehat{\boldsymbol{\nu}}_{\mathrm{PQL}})^T\boldsymbol{\Sigma}^{-1}(\boldsymbol{\nu} - \widehat{\boldsymbol{\nu}}_{\mathrm{PQL}})
$$

$$
- \sum_{\ell=1}^{L}\big((A_{u\ell} + \frac{q_\ell}{2} + 1)\log\sigma_{u\ell}^2 + (A_{u\ell} + \frac{1}{2}\|\mathbf{u}_\ell\|^2)/\sigma_{u\ell}^2\big)\big\}\Big]
$$

$$
= \exp\Big[\gamma_s\big\{\mathbf{y}^T\mathbf{C}\boldsymbol{\nu} - \mathbf{1}^T b(\mathbf{C}\boldsymbol{\nu}) - \frac{1}{2\sigma_\beta^2}\|\boldsymbol{\beta}\|^2\big\} + \sum_{\ell=1}^{L}(A_{u\ell} + \frac{q_\ell}{2})\log(A_{u\ell} + \frac{1}{2}\|\mathbf{u}_\ell\|^2)
$$

$$
+ (1 - \gamma_s)\big\{-\frac{1}{2}(\boldsymbol{\nu} - \widehat{\boldsymbol{\nu}}_{\mathrm{PQL}})^T\boldsymbol{\Sigma}^{-1}(\boldsymbol{\nu} - \widehat{\boldsymbol{\nu}}_{\mathrm{PQL}})\big\} \tag{11}
$$

$$
- \sum_{\ell=1}^{L}\big\{(A_{u\ell} + \frac{q_\ell}{2} + 1)\log\sigma_{u\ell}^2 + (A_{u\ell} + \frac{1}{2}\|\mathbf{u}_\ell\|^2)/\sigma_{u\ell}^2\big\}\Big]
$$

In the absence of any additional information about the shapes of these distributions, it is difficult to specify a sensible generic sequence of $\gamma_s$ values. Hence for the rest of the paper we choose to increase $\gamma_s$ from $\gamma_0 = 0$ to $\gamma_{S-5} = 1$ in a linear fashion, that is, values of $\gamma_s$ are sequentially incremented by the same amount. Additionally, we append $\gamma_{S-4} = \cdots = \gamma_S = 1$ to this sequence to give five stages at the end of the sampler on which the particles are not resampled. This means that the final sample is well spread out over the distribution $\pi$ (it was found that if resampling happened too close to the end of the sampler then several samples might be identical, resulting in poor density estimates being produced using the standard techniques).

It is an interesting and open research question as to whether the sequence $\gamma_s$ can be chosen in a more principled manner. One option would be to choose the sequence in advance using some properties of the distributions $\pi_0$ and $\pi$. An alternative would be to choose the next $\gamma_s$ adaptively while the sampler proceeds through the sequence of distributions; however it is not straightforward to generalise the proofs of validity of the sampler in this case.

### *3.3. Transition kernels*

For this paper we choose to use Metropolis-Hastings transition kernels for the parameters in $\boldsymbol{\nu}$. The choice of inverse gamma distributions for the components of $\boldsymbol{\sigma}^2$ within $\pi_0$ means that we can simply use Gibbs sampling steps, [17], to update those components. At each step $s$ we use a Metropolis–Hastings transition kernels $K_s$. Since $\pi_0$ is an approximation to $\pi$, and $\pi_s$ is in some sense between $\pi_0$ and $\pi$, we use the same proposal distributions at each step $s$. These proposal distributions are derived from $\pi_0$ as described in this section.

We form a partition $\{\mathcal{I}_1, \ldots, \mathcal{I}_J\}$ of $\{1, \ldots, P\}$, where $P$ is the number of columns in $\mathbf{C}$, so that $[\mathbf{C}_{\mathcal{I}_1} \cdots \mathbf{C}_{\mathcal{I}_J}]$ is the matrix $\mathbf{C}$, but with columns possibly re-ordered; and

$$\begin{bmatrix} \boldsymbol{\nu}_{\mathcal{I}_1} \\ \vdots \\ \boldsymbol{\nu}_{\mathcal{I}_J} \end{bmatrix}$$

is the corresponding partition of $\boldsymbol{\nu}$. (The case $J = 1$ corresponds to no partitioning.) On each move step of the algorithm we move through the series of subsets $\mathcal{I}_j$, for $j = 1, \ldots, J$. We apply a Metropolis-Hastings transition kernel to the components $\boldsymbol{\nu}_{\mathcal{I}_j} = (\nu_i)_{i \in \mathcal{I}_j}$.

To describe the transitions we introduce the matrices $\boldsymbol{\Sigma}_{\mathcal{I}_j}$, where $\boldsymbol{\Sigma}_{\mathcal{I}_j}$ is the conditional covariance under $\pi_0$ of $\boldsymbol{\nu}_{\mathcal{I}_j}$ given the values of $\boldsymbol{\nu}_{-\mathcal{I}_j} = (\nu_i)_{i \notin \mathcal{I}_j}$. These can be calculated at the start of the algorithm. Recall that since $\pi_0$ is an approximation of $\pi$, the $\boldsymbol{\Sigma}_{\mathcal{I}_j}$ matrices therefore correspond to approximations of the conditional covariance of $\boldsymbol{\nu}_{\mathcal{I}_j}$ given $\boldsymbol{\nu}_{-\mathcal{I}_j}$ under the posterior distribution $\pi$. Note that we are here assuming that the approximate covariance matrices $\boldsymbol{\Sigma}_{\mathcal{I}_j}$ are close enough to the truth to be useful as proposal distributions in the random walk Metropolis–Hastings kernel.

The proposal distribution for $\boldsymbol{\nu}_{\mathcal{I}_j}$ is then a normal distribution centered on the current value of $\boldsymbol{\nu}_{\mathcal{I}_j}$ with covariance $\tau_j^{\nu} \boldsymbol{\Sigma}_{\mathcal{I}_j}$. The acceptance probability for the move, applied after reweighting to get a weighted distribution from $\pi_s$, is simply calculated from the ratio of $\pi_s$ values for the proposed and current values.

The scaling parameters $\tau_j^{\nu}$ are by default chosen to be $2.4/\sqrt{|\mathcal{I}_j|}$ following the heuristic of [25]. However in practice they are usually chosen, based on several runs of the algorithm, to ensure that the acceptance rates remain close to 0.23 (again following [25]). Details of specific choices used are given in the examples.

To update the variance parameters $\boldsymbol{\sigma}^2$, a Gibbs sampling step can be applied. Note from (11) that for each $s$ the full conditional distribution of $\sigma_{u\ell}^2$ is simply an inverse gamma distribution, depending on the corresponding vector of regression coefficients $\mathbf{u}_\ell$. However if a different prior is used for $\boldsymbol{\sigma}^2$ then Gibbs sampling will not be available and a Metropolis–Hastings update should be performed for each $\sigma_{u\ell}^2$ in turn.

## 4. Examples

In this section we will demonstrate the methodology on two examples. The first example is a semiparametric Poisson regression model, with simulated data so that fair comparisons can be drawn with alternative MCMC approaches. The second example is a binary logistic regression involving respiratory infection in Indonesian children, with both a semiparametric component and random effects. All computations were carried out in the R language [29], using a single core AMD Opteron 2.0GHz, this is similar to running the program on a standard PC.

### *4.1. Semiparametric Poisson regression*

In this section, we generate $n = 500$ Poisson random variables $y_i, i = 1, \ldots, n$ from

$$y_i \sim \text{Poisson}(\exp\{0.7x_{1i} + 2x_{2i} + \cos(4\pi x_{2i})\})$$

where $x_{1i}$ is 0 or 1 with probability 0.5, and $x_{2i}$ is uniformly sampled from the interval $[0, 1]$.

We fit model (3), with

$$b(x) = e^x, \qquad \boldsymbol{\beta} = \begin{bmatrix} \beta_0 \\ \beta_1 \\ \beta_2 \end{bmatrix}, \qquad \mathbf{X} = \begin{bmatrix} 1 & x_{11} & x_{21} \\ 1 & x_{12} & x_{22} \\ \vdots & \vdots & \vdots \\ 1 & x_{1n} & x_{2n} \end{bmatrix}.$$

The radial cubic basis function is used to model the function $f(x_{2i}) = \cos(4\pi x_{2i})$. This implies modelling $f(x_{2i}) = \beta_{x_2} x_{2i} + \mathbf{Z}_{x_{2i}} \mathbf{u}$, where for knot points $\kappa_k$, chosen to be the $(\frac{k+1}{K+2})$th quantile of the unique predictor values, for $k = 1, \ldots, K, K = 10$,

$$\mathbf{u} = \begin{bmatrix} u_1 \\ \vdots \\ u_{10} \end{bmatrix}, \quad [\mathbf{u}|\sigma_u^2] \sim N(\mathbf{0}, \sigma_u^2 \mathbf{I}), \quad \text{and} \quad \mathbf{Z}_{x_{2i}} = [|x_{2i} - \kappa_k|^3]_{1 \le k \le 10} [|\kappa_{k'} - \kappa_k|^3]_{1 \le k, k' \le 10}^{-1/2}$$

The `glmmPQL` method of the R statistical package gives an approximate MLE for the regression coefficients $\widehat{\boldsymbol{\nu}}_{\text{PQL}}$, and the variance parameters $\widehat{\boldsymbol{\sigma}}_{\text{PQL}}^2$. We follow the general algorithm given in Section 3. There are 13 regression coefficients to be estimated for this model, and one variance parameter. We updated each of the regression coefficients $\boldsymbol{\nu}$ singly using random walk Metropolis-Hastings (RWMH) updates for the move step of the algorithm and a Gibbs update for the variance parameter. As with MCMC, the tuning of this kernel is crucial to the success of the algorithm; to achieve an acceptance rate in the MCMC step between 20–30% we set $\tau_{\mathcal{I}}^{\nu} = 1/3$. We also choose the number of steps $S = 105$ and the number of particles $N = 1000$ based on preliminary runs. We will discuss the choices of $N$ and $S$ in more detail in Section 5.

We compare the performance of the SMC sampler simulations by monitoring the QQ-plots of samples from a simple importance sampler, a single-variable slice sampler which updates one parameter at a time and a standard RWMH sampler with the same transition kernels as used in the SMC sampler (i.e. those described in Section 3.3). Figure 1(a) shows the QQ-plot for the $\beta_1$ parameter, and the corresponding density estimates for $\beta_1$ is given in (b). With the exception of the importance sampler, which can perform badly on different simulated data sets, the remaining samplers achieved good concordance. This required 1,000 particles with 105 steps for the SMC sampler. For comparison, we used 20,000 iterations of both slice sampler and RWMH MCMC scheme, with the first 10,000 discarded as burn-in. These took approximately 1394 and 2430 seconds respectively, whereas the SMC sampler took approximately 700 seconds. The majority of the gains in computational time for the SMC sampler come from the fact that the particles at each step can be updated simultaneously without the need to cycle through a loop, compared with MCMC samplers where each iteration has to be updated sequentially depending on the value of the parameter at the previous step. In the R programming language used for this research, as with many other high level programming languages, this provides a very significant computational advantage.

The nonparametric fits of the model, calculated using the estimated posterior mean of $\mathbf{u}$, are displayed in Figure 2. The model has successfully recovered the nonlinearity in the dependency on $x_2$ and fits the data well.

### 4.2. Example: Respiratory infection in Indonesian children

Here we apply sequential Monte Carlo algorithm to an example involving respiratory infection in Indonesian children (see [10, 22]). The data contain longitudinal measurements on 275 Indonesian children, where the indicator for respiratory infection is the binary response. The covariates include age, height, indicators for vitamin A deficiency, sex, stunting and visit numbers (one to six).

Previous analyses have shown the effect of age of the child to be non-linear, hence we use a logistic additive mixed model of the form

$$\mathrm{logit}\{P(\mathrm{respiratory\ infection}_{ij} = 1|U_i, u_k)\} = \beta_0 + U_i + \boldsymbol{\beta}^T x_{ij} + f(\mathrm{age}_{ij})$$

for $1 \leq i \leq 275$ children and $1 \leq j \leq n_i$ repeated measures within a child. $U_i \overset{\mathrm{i.i.d.}}{\sim} N(0, \sigma_U^2)$ is a random child effect, $x_{ij}$ is the measurement on a vector of the remaining 9 covariates, and $f$ is modelled using penalized splines with spline basis coefficients $u_k$ i.i.d. $N(0, \sigma_u^2)$.

As recommended by [16], we use hierarchical centering of random effects. All continuous covariates are standardised to have zero mean and unit standard deviation, so that the choices of hyperparameters can be independent of scale. Radial cubic basis functions are used to fit the covariate age, where

$$f(\mathrm{age}) = \beta_{\mathrm{age}}\mathrm{age} + \mathbf{Z}_{\mathrm{age}}\mathbf{u}$$

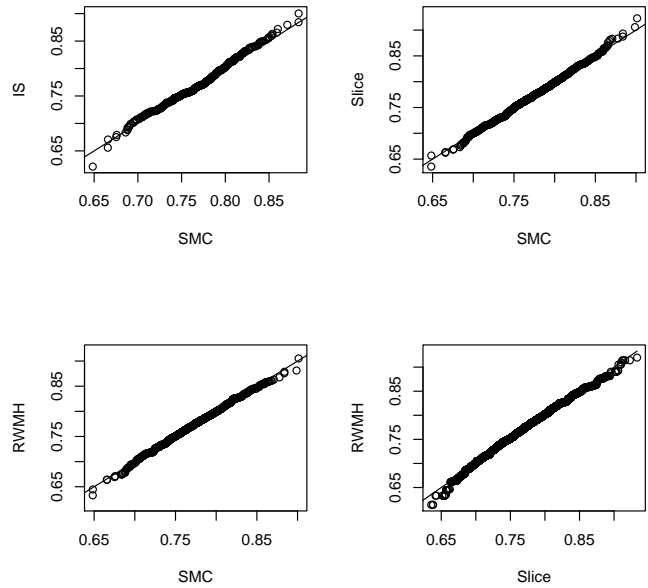

(a)

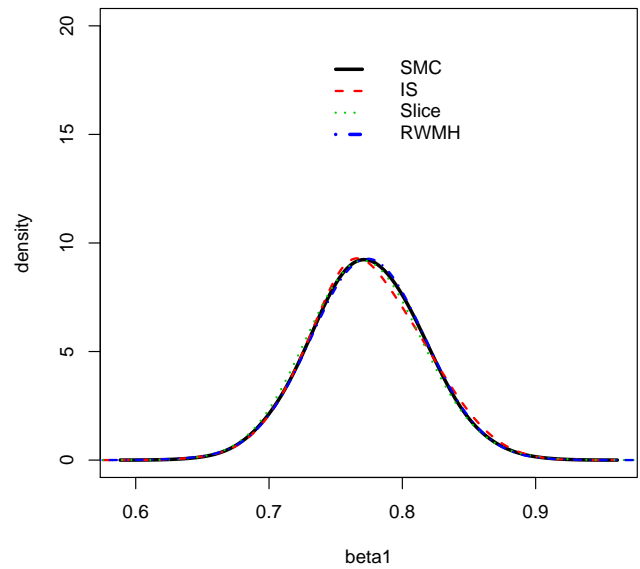

(b)

FIG 1. *QQ-plots of SMC sampler output against simple importance sampler, the slice sampler and the RW Metropolis-Hastings sampler for $\beta_1$ (a). The corresponding density estimates (b).*

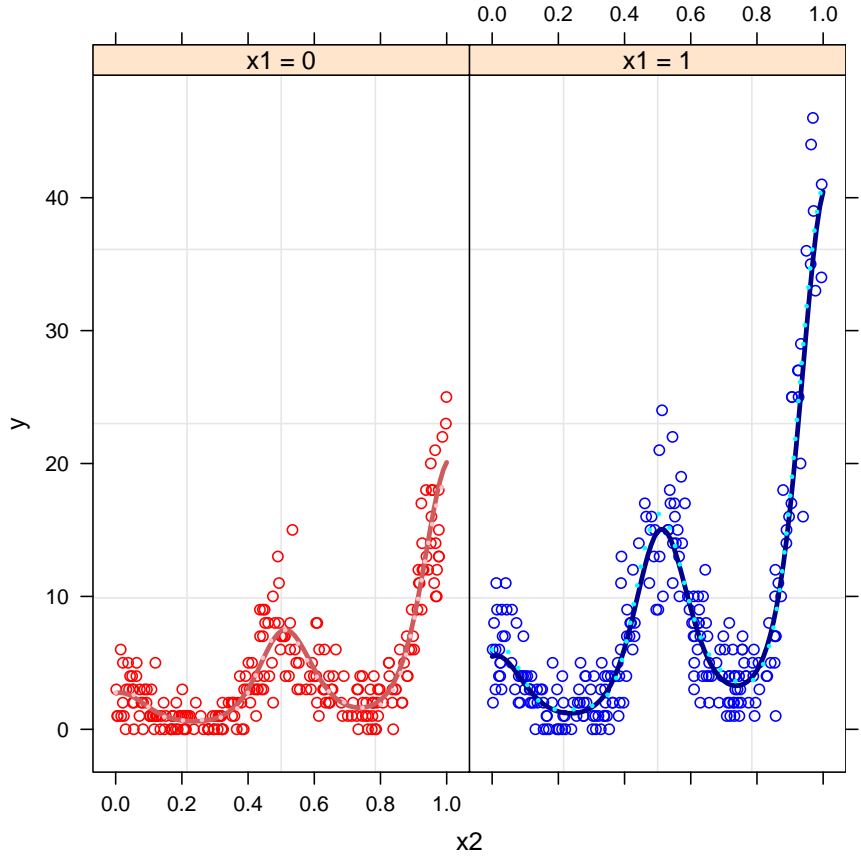

FIG 2. *The data, the true mean values (solid lines), and the estimated mean values (dotted lines) for the simulated Poisson example. The fit was based on a SMC sampler run with 1000 particles and 105 intermediate steps, as described in the text.*

where

$$\mathbf{Z}_{\mathrm{age}} = [|\mathrm{age} - \kappa_k|^3]_{1 \le \kappa \le K} [|\kappa_k - \kappa_{k'}|^3]_{1 \le k, k' \le K}^{-1/2} \quad \text{and} \quad \mathbf{u} \sim N(0, \sigma_u^2 \mathbf{I})$$

with $\kappa_k$ chosen to be the $(\frac{k+1}{K+2})$th quantile of the unique predictor values. We take $K = 20$ in this example.

We use a vague prior $N(0, 10^8)$ for the fixed effects. For both variance components, we use the conjugate Inverse Gamma prior IG(0.01, 0.01). Other prior choices are available, see [30]. Here, random walk Metropolis-hastings updates were carried out for each regression coefficients separately, with Gibbs sampling used for the variance parameters. The tuning parameters $\tau_{\mathcal{I}_j}^{\nu}$ for the Metropolis-Hastings update are again chosen to achieve an acceptance rate in the MCMC

| coeff. | density | summary |
|--------|---------|---------|
| vit. A defic. | 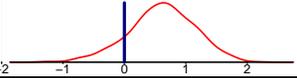 | posterior mean: 0.61<br>95% credible interval:<br>(−0.542,1.62) |
| male | 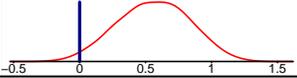 | posterior mean: 0.563<br>95% credible interval:<br>(0.0439,1.06) |
| height | 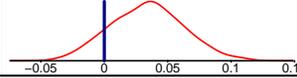 | posterior mean: 0.0338<br>95% credible interval:<br>(−0.0208,0.0893) |
| stunted | 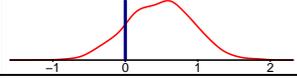 | posterior mean: 0.474<br>95% credible interval:<br>(−0.402,1.31) |
| visit 2 | 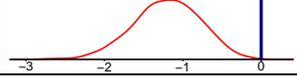 | posterior mean: −1.2<br>95% credible interval:<br>(−2.1,−0.431) |
| visit 3 | 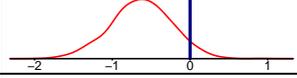 | posterior mean: −0.629<br>95% credible interval:<br>(−1.41,0.11) |
| visit 4 | 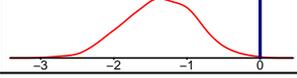 | posterior mean: −1.37<br>95% credible interval:<br>(−2.3,−0.467) |
| visit 5 | 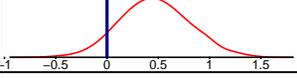 | posterior mean: 0.468<br>95% credible interval:<br>(−0.158,1.14) |
| visit 6 | 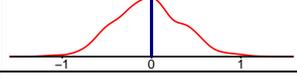 | posterior mean: −0.0384<br>95% credible interval:<br>(−0.722,0.67) |
| st.dev.(subject) | 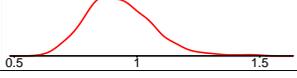 | posterior mean: 0.928<br>95% credible interval:<br>(0.7,1.23) |

Fɪɢ 3. *Summary of coefficients in respiratory infections in Indonesian children example.*

step between 20–30%, we used $\tau^{\nu}_{\mathcal{I}_j} = 3$ for the fixed effect coefficients, $\tau^{\nu}_{\mathcal{I}_j} = 6$ for the random effect coefficients, and $\tau^{\nu}_{\mathcal{I}_j} = 5$ for the spline coefficients.

Figure 3 show the results from simulation, using 1000 particles and 305 intermediate steps. The Figure shows borderline positive effect of Vitamin A deficiency, sex and some visit numbers on respiratory infection. These results are in keeping with previous analyses. Figure 4(a) shows the nonlinear effect of age; 4(b) shows the effective sample size at each of 300 sequential steps of the simulation, vertical lines indicate the occurrence of resampling.

Again, we compare the performance of the SMC sampler with the importance sampler, slice sampler and RWMH sampler with the same transition kernel as Step 3 of the SMC sampler. Results for 5,000 samples of the importance sampler, 1,000 SMC particles and 5,000 slice samples with 5,000 burn-in and 5,000 RWMH samples with 5,000 burn-in are plotted in Figure 5, good agreements

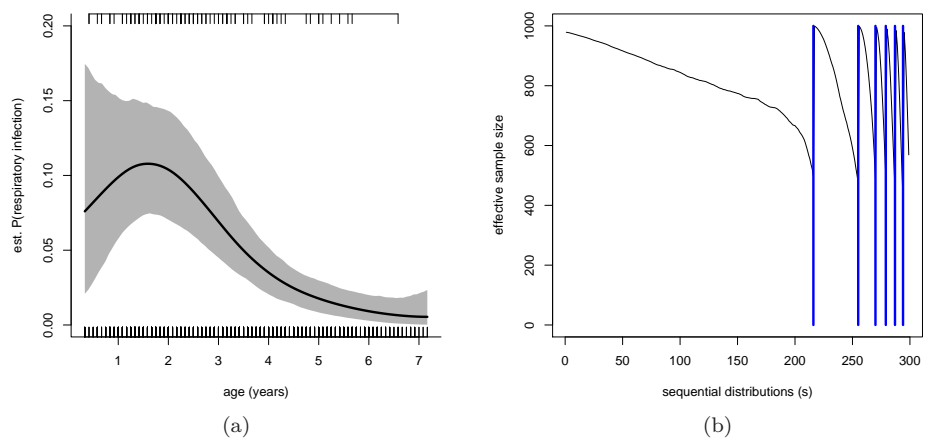

(a)                         (b)

Fɪɢ 4. *Respiratory infections in Indonesian children example. (a) Posterior mean of the estimated probability of respiratory infection* $f(\texttt{age})$ *with all other covariates set to their average values. (b) Effective sample size over 300 distributions, vertical lines indicate instances of resampling.*

are found between the SMC, slice and MCMC samplers. The sampler which performed badly appears to be the importance sampler, where in this case, the sampler appears to suffers from particle depletion where one or few particles from an area of high posterior density is dominating the other particles.

Finally, the SMC sampler took approximately the same amount of time as the slice sampler at 2.8 hours and the RWMH took about 9 hours (similarly 9 hours was required in `WinBUGS`). In the next section we discuss the effect of the sample size and step size specifications on the efficiency of the SMC sampler.

## 5. Improving sampler performance

In this section we investigate the SMC sampler performance by looking at the effects of user-defined specifications such as the number of sequential steps ($S$); the number of particles to sample ($N$) and block updating strategies.

Here we base efficiency comparisons on the effective sample size diagnostic calculation of [4]. (Note this is different from the ESS [20] used in determining whether resampling is performed.) This diagnostic is essentially an analysis of variance approach based on several parallel runs of the algorithm, which provides the number of independent samples from the posterior distribution that would be required to gain the same degree of accuracy: higher numbers are obviously preferrable. This estimate of effective sample size is not affected by resampling and in addition, can be used to compare SMC sampler with MCMC approaches under a consistent framework. Thus, for a given parameter, we calculate the

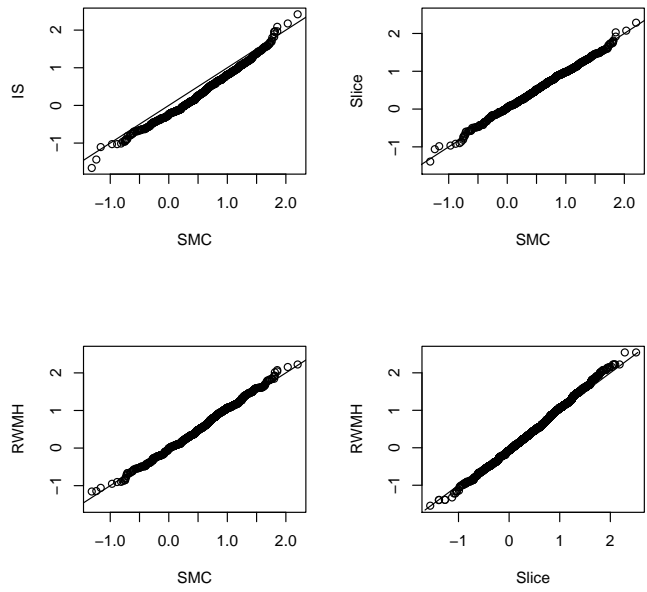

(a)

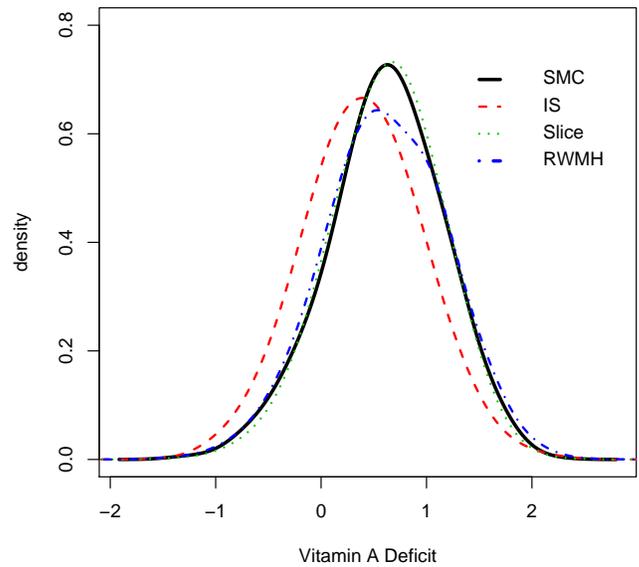

(b)

Fig 5. *QQ-plots of SMC sampler output against simple importance sampler, the slice sampler and the RW Metropolis-Hastings sampler for the coefficient of vitamin A deficiency (a). The corresponding density estimates (b).*

### Algorithm efficiency for PMM example

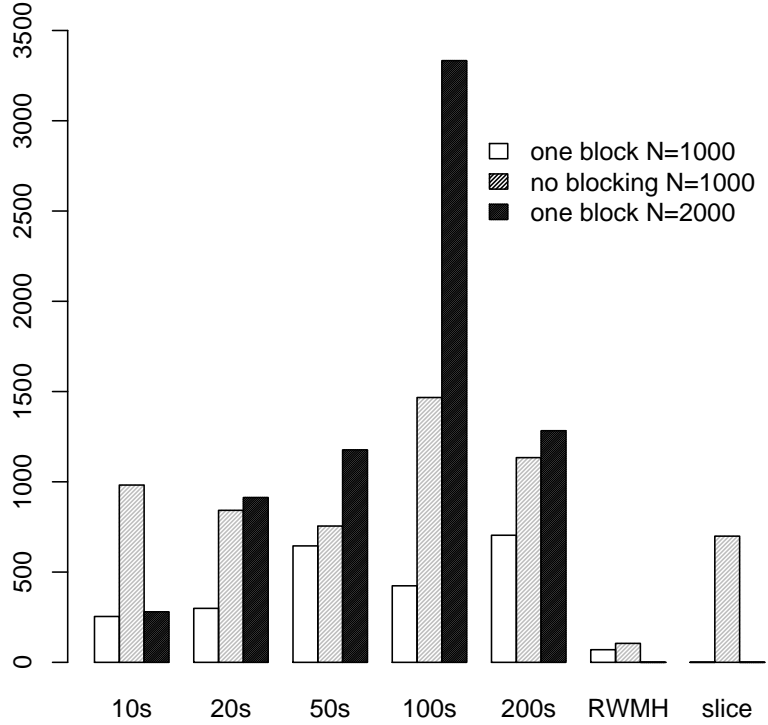

FIG 6. *Comparison of effective sample size for $\beta_1$ from the SMC sampler (over increasing number of sequential distributions (S), and the slice and RWMH samplers in the Poisson regression examples.*

ratio of the average estimate of the posterior variance of the parameter to the variance of the posterior means of the parameter across independent runs. All our experiments were carried out using the two examples in Sections 4.1 and 4.2, where $\beta_1$ is used for the Poisson regression example, and the coefficient of Vitamin A deficiency is used for the logistic regression example.

In experimenting with block/simultaneous updating the parameters for the move step of the SMC sampler, we found that in some cases, particularly for the logistic example, naïve blocking updating (i.e., blocking regression coefficients, random effects coefficients, and spline coefficients) can in fact adversely affect the performance of the sampler. We also found that careful tuning of acceptance probabilities in the RWMH step to be between 20-30% can be crucial to the performance. Finally, we found that increasing the number of sequential distributions $S$, as well as increasing the number of particles $N$ can greatly improve the effective sample size.

Figure 6 shows the effective sample size for $\beta_1$ calculated from a total of 10 independent runs of the sampler for the Poisson regression example. We calculated the diagnostic for the SMC sampler over $S = 10, 20, 50, 100, 200$. For each $S$, we implemented the sampler with a single block update using $N = 1000$ and $N = 2000$ particles, and a sampler without blocking with $N = 1000$. For comparison, the diagnostic for $\beta_1$ was also calculated for a (single-variable) slice sampler and a RWMH sampler, each of 20,000 iterations with 10,000 burnin. The RWMH algorithm used the same transition kernels as Step 3 of the SMC sampler, with and without block updating.

Clearly, the effective sample size of the sampler with no blocking is larger than that of the sampler with one block for all $S$, and increases with larger values of $S$ and $N$. We can see that the effective sample size of the slice sampler is comparable to the SMC sampler with $S = 50$ and $N = 1000$, while the RWMH sampler achieves the smallest effective sample size.

For the logistic regression example of Section 4.2, we found that by updating regression parameters singly, and choosing $S = 200$ and $N = 2000$ achieved an effective sample size of 2604, and $S = 300$ and $N = 1000$ achieved an effective sample size of 2377. The two samplers took about 3.8 and 2.9 hours to run respectively. As a comparison, the slice sampler with 10,000 iterations with 5,000 burn in took 2.8 hours to run and achieves an effective sample size of 1948, while a RWMH sampler of length 10,000 with 5,000 burn in achieves an effective sample size of only 177, and taking 9 hours to run.

## 6. Conclusion

In this paper we presented a general sequential Monte Carlo algorithm to produce samples from the posterior distribution for Bayesian analysis of generalised linear mixed models. The algorithm is an alternative to the popular Markov chain Monte Carlo methods. We have demonstrated that the algorithm can handle problems where the number of parameters to be estimated in the model is high. For example, in the spline formulation of the Indonesian children example, there were over 300 parameters in the model. In addition, the algorithm is generally easy to apply. We have also demonstrated that it can have substantial gains in computational time over traditional MCMC in both a simulated poisson example and a real data binomial example. Finally, perhaps the biggest advantage of SMC over MCMC samplers is the fact convergence of SMC samplers does not rely on convergence of Markov chains, which can be problematic in designing more efficient algorithms in complex problems.

We have found that in the context of Bayesian GLMM analysis, the design of the initial sequential Monte Carlo distribution may be helped by using approximate parameter estimates from classical GLMM analysis, such as using the PQL method to find MLE of the likelihood. Note that in the case where such estimates cannot be easily found, and the only sensible choice is a diffuse prior, then a SMC sampler with many more particles and sequential distributions will be needed to obtain good results. In choosing the schedule for the tempering sequence $\gamma_s$ in (7), we have found no substantial difference between the different types of schedules currently used in the literature, hence we recommend that a simple linear schedule be adopted in the GLMM context. We have also found that by tuning the acceptance rate of the Random-Walk Metropolis-Hastings kernel in the Move step of the SMC sampler to around 20-30% significantly improves the performance of the sampler, see [9], although this practical finding does not yet have rigorous theoretical support.

Finally, in implementing the Move step of the SMC sampler, one has some degree of flexibility when the Markov chain Monte Carlo update is used. For example, one may consider a better choice of proposal distributions for the Metropolis-Hastings algorithm, by allowing the algorithm to automatically scale a proposal distribution, see for example [5]. Here a major advantage over the traditional MCMC is that the algorithm does not suffer from the restrictions associated with a Markov chain, and information from previous samples can be freely used to obtain future samples. Furthermore, one is not restricted to only MCMC type of moves in this step, other move types are possible, see [8].

However, sequential Monte Carlo algorithms are not black-box algorithms, requiring a certain amount of tuning and user input. In particular, one needs to set the number of sequential distributions ($S$) the number of particles to sample ($N$) and tuning parameters for the Metropolis-Hastings kernels in the move step of the algorithm.

### Acknowledgements

This research was partially supported by Australian Research Council Discovery Project DP0877432 (Y. Fan), Nuffield Foundation grant number NAL/00803/G (D.S. Leslie) and Australian Research Council Discovery Project DP0877055 (M.P. Wand).

### Appendix: Algorithmic description of the SMC sampler method for GLMMs

In this appendix we give a detailed description of how to use the SMC sampler method to perform inference in GLMMs. We use the notation of Section 2; choices made in the implementation of the algorithm are explained in Section 3.

For any subset $\mathcal{I}$ of $\{1, \ldots, P\}$ we write $\mathbf{C}_{\mathcal{I}}$ for the submatrix of of the design matrix $\mathbf{C}$ consisting of columns in $\mathcal{I}$, $\mathbf{C}_{-\mathcal{I}}$ for the submatrix consisting of columns of $\mathbf{C}$ not in $\mathcal{I}$, $\boldsymbol{\nu}_{\mathcal{I}}$ and $\boldsymbol{\nu}_{-\mathcal{I}}$ for the analogously defined subvectors of $\boldsymbol{\nu}$. Also for any square matrix $\mathbf{Q}$ we write $\mathbf{Q}_{\mathcal{I}\mathcal{I}}$ for the square submatrix corresponding to rows and columns in $\mathcal{I}$, $\mathbf{Q}_{\mathcal{I},-\mathcal{I}}$ for the submatrix with rows in $\mathcal{I}$ and columns not in $\mathcal{I}$, $\mathbf{Q}_{-\mathcal{I},\mathcal{I}}$ for the submatrix with rows not in $\mathcal{I}$ and columns in $\mathcal{I}$, and $\mathbf{Q}_{-\mathcal{I},-\mathcal{I}}$ for the square submatrix with rows and columns not in $\mathcal{I}$.

### *Initialisation*

- Set the number of particles $N$ and the number of intermediary distributions $S$.
- Construct a vector $\boldsymbol{\gamma}$ with $s$th entry $\psi(s)$, $s = 0, 1, \ldots, S$, where $\psi : \{0, 1, \ldots, S\} \to [0, 1]$ is an increasing function such that $\psi(0) = 0$ and $\psi(S) = 1$. For the results in this paper we used $\psi(s) = \min\{1, s/(S-5)\}$.
- Construct subsets $\mathcal{I}_1, \ldots, \mathcal{I}_J$ of $\{1, \ldots, P\}$ such that $\bigcup_{j=1}^{J} \mathcal{I}_j = \{1, \ldots, P\}$. The case $J = 1$ corresponds to no blocking of variables for the move step.
- Set tuning parameters $\tau_j^\nu > 0$, $j = 1, \ldots, J$ for the Metropolis–Hastings updates. Usually these will be set based on preliminary runs of the algorithm, and convenient defaults are $\tau_j^\nu = 2.4/\sqrt{|\mathcal{I}_j|}$.
- Use the Breslow & Clayton (1993) penalised quasi-likelihood (PQL) algorithm to obtain initial estimates:

$$\widehat{\boldsymbol{\nu}}_{\text{PQL}} \quad \text{and} \quad \widehat{\boldsymbol{\sigma}}_{\text{PQL}}^2.$$

  This is facilitated by software such as `glmmPQL()` in the R package `MASS` [29]. Use these estimates in (8) to calculate $\boldsymbol{\Sigma}$.
- For each $j = 1, \ldots, J$, calculate the conditional covariance under $\pi_0$ of $\boldsymbol{\nu}_{\mathcal{I}}$ conditional of $\boldsymbol{\nu}_{-\mathcal{I}}$. If $Q = \Sigma^{-1}$, then this conditional covariance is $\boldsymbol{\Sigma}_{\mathcal{I}_j} := (Q_{\mathcal{I}\mathcal{I}})^{-1}$.

### *Initial sample from $\pi_0$*

- Produce a sample of size $N$ from $\pi_0$: for each $i = 1, \ldots, N$ sample $\nu_i$ from the normal distribution with mean $\widehat{\boldsymbol{\nu}}_{\text{PQL}}$ and covariance $\boldsymbol{\Sigma}$, then sample $\boldsymbol{\sigma}_i^2$ from the conditional inverse gamma distributions (9).
- Set the weights $w_i = 1/N$ for each $i = 1, \ldots, N$.

### *Sequential sampling from each $\pi_s$*

For each $s = 1, \ldots, S$ in turn,

**Reweight** For each $i = 1, \ldots, N$, update $w_i$ according to

$$w_i \leftarrow w_i \frac{\pi_s(\nu_i, \boldsymbol{\sigma}_i^2)}{\pi_{s-1}(\nu_i, \boldsymbol{\sigma}_i^2)} = \left( \frac{\pi(\nu_i, \boldsymbol{\sigma}_i^2)}{\pi_0(\nu_i, \boldsymbol{\sigma}_i^2)} \right)^{\gamma_s - \gamma_{s-1}}$$

then normalise the weights by setting $w_i \leftarrow w_i / \sum_{j=1}^{N} w_j$. To avoid overflow and underflow problems it is recommended that logarithms be used in this step.

**Resample** Calculate the effective sample size (ESS) using

$$ESS = (\sum_{i=1}^{N} w_i)^2 / \sum_{i=1}^{N} (w_i)^2.$$

If $ESS < N/2$ (or if $s = \min\{s : \gamma_s = 1\}$) then resample the particles. The naive version of resampling, which introduces unnecessary Monte Carlo variation into the scheme, simply samples (with replacement) from the pool of particles, with particle $i$ selected with probability $w_i$. However in our implementation we use stratified resampling [21] to reduce the Monte Carlo variation. After resampling set $w_i = 1/N$ for all $i = 1, \dots, N$.

**Move**
- For each $j = 1, \dots, J$ and each $i = 1, \dots, N$, generate proposals $(\tilde{\boldsymbol{\nu}}_i)_{\mathcal{I}_j} \sim N((\tilde{\boldsymbol{\nu}}_i)_{\mathcal{I}_j}, \tau_j^\nu \Sigma_{\mathcal{I}_j})$, $1 \le i \le N$. With probability

$$\alpha^{(i)} = \max\left\{1, \frac{\pi_s((\tilde{\boldsymbol{\nu}}_i)_{\mathcal{I}_j} | (\boldsymbol{\nu}_i)_{-\mathcal{I}_j}, \boldsymbol{\sigma}_i^2)}{\pi_s(\boldsymbol{\nu}_i, \boldsymbol{\sigma}_i^2)}\right\}$$

accept the proposal and set $(\boldsymbol{\nu}_i)_{\mathcal{I}_j} = (\tilde{\boldsymbol{\nu}}_i)_{\mathcal{I}_j}$. Otherwise reject the proposal and leave $(\boldsymbol{\nu}_i)_{\mathcal{I}_j}$ unchanged. Again, it is recommended that logarithms be used when calculating $\alpha$ to avoid overflow and underflow problems. Note that several parts of the ratio in the calculation of $\alpha$ are the same in both the numerator and denominator and need not be calculated.

- For each $\ell = 1, \dots, L$, and for each $i = 1, \dots, N$, sample $(\boldsymbol{\sigma}_i^2)_\ell$ from the inverse gamma distribution with shape $A_{u\ell} + q_\ell/2$ and rate $A_{u\ell} + \|\mathbf{u}_\ell\|^2/2$. Note that if inverse gamma distributions are not used as the prior distribution for $\boldsymbol{\sigma}^2$ then sampling from inverse gamma distributions here would not result in a transition kernel that admits $\pi_s$ as a stationary distribution. Instead further Metropolis–Hastings can be used for each $\sigma_{u\ell}^2$ in turn.

Note that the decision to resample on the first step at which $\gamma_s = 1$ means that the final sample is an unweighted sample from $\pi$. Hence standard techniques for dealing with samples from posterior distributions can be used. However the plug-in rule for the bandwidth used in the density estimates performed poorly for resampling close to step S, since some particles were identical. This is the reason that we generally set $\gamma_{S-5} = 1$ and finish with five applications of the transition kernel to the unweighted sample, resulting in a suitably diverse sample from $\pi$.